\documentclass[twocolumn,showpacs,prl]{revtex4}

\usepackage{graphicx}
\usepackage{dcolumn}
\usepackage{bm}
\usepackage{subfigure}
\usepackage{amssymb,amsmath}

\begin{document}


\title{Modeling elastic instabilities in nematic elastomers}%

\author{Badel L. Mbanga}
\author{Fangfu Ye}
\altaffiliation[Current address: ]{Dept. of Physics, Univ. of Illinois at Urbana-Champaign, Urbana, IL 61801.}
\author{Jonathan V. Selinger}%
\author{Robin L. B. Selinger}
 \email{rselinge@kent.edu}
\affiliation{
Liquid Crystal Institute, Kent State University, Kent, Ohio 44242\\
}

\date{\today}

\begin{abstract}

Liquid crystal elastomers are cross-linked polymer networks covalently bonded
with liquid crystal mesogens. In the nematic phase, due to strong coupling
between mechanical strain and orientational order, these materials display
strain-induced instabilities associated with formation and evolution of
orientational domains. Using a 3-d finite element elastodynamics simulation, 
we investigate one such instability, the onset of stripe formation in
a monodomain film stretched along an axis perpendicular to the nematic director.
In our simulation we observe the formation of striped domains with alternating director
rotation. This model allows us to explore the fundamental physics governing dynamic mechanical 
response of nematic elastomers and also provides 
a potentially useful computational tool for engineering device applications.
\end{abstract}

\maketitle


Liquid crystal elastomers (LCE) exhibit some of the elastic properties of
rubber along with the orientational order properties of liquid crystals,
displaying a variety of nematic and smectic phases. They
are composed of liquid crystal mesogens covalently bonded to a cross-linked
polymer backbone \cite{Warnerbook,Brand_review}. These materials display strong coupling
between orientational order of the mesogens and mechanical deformation of the
polymer network. For instance in a nematic LCE, any change in the magnitude of
the nematic order parameter induces shape change, e.g. the isotropic-nematic
phase transition induces strains of up to several hundred percent \cite{Jawad}.

Conversely, applied strain can also drive changes in orientational order,
producing the fascinating phenomenon of semisoft elasticity \cite{fangfu_prl}.
In a classic experiment, 
Kundler and Finkelmann \cite{kundler_finkelmann} measured the
mechanical response of a monodomain nematic LCE thin film stretched along an axis
perpendicular to the nematic director. They observed a semi-soft elastic
response with a pronounced plateau in the stress-strain curve arising at a
threshold stress. Accompanying this instability they observed the formation of
striped orientational domains with alternating sense of director rotation, and
a stripe width of 15 $\mu$m. They repeated the experiment with samples cut at
different orientations to the director axis, and found that the instability was
absent when the angle between the initial director and the stretch axis
was less than $70^o$ ; in this geometry, instead of forming stripes, the director rotates
smoothly as a single domain. 

DeSimone et al \cite{desimone} carried out numerical simulation studies of the
stripe instability using a two-dimensional finite element elastostatic method. 
Each area element in the system was considered as a composite of domains with 
different orientations. This simulation model was the first to reproduce successfully the
soft elastic response of nematic elastomers, but did not attempt to resolve the resulting
microstructural evolution.

Here we explore this elastic instability in more detail by simultaneously modeling the sample's
mechanical response and the associated microstructural evolution as a function of strain. 
We use a Hamiltonian-based 3-d finite element elastodynamics model 
with terms that explicitly couple strain and nematic order. 
By resolving the finite element  mesh down to the micron scale, we resolve 
the formation of orientational domains, and because
the model is dynamic rather than static in character, we can examine the effects
of strain rate. We use the simulation to explore the dependence of mechanical
response on deformation geometry.

We model this instability in a thin film of nematic elastomer which has been cross-linked in the nematic phase. Using public domain meshing software \cite{salome} we discretize the volume of the sample into approximately $78,000$ tetrahedral elements. For each volume element we assign a local variable $\bf n$ 
which defines the nematic director, and $Q_{ij}=\frac{1}{2}S(3n_{i}n_{j}-\delta_{ij})$ which is the associated
symmetric and traceless nematic order tensor. The initial state is taken to be a monodomain with
$\bf n= \bf n_o$ in every element; this configuration is defined as the system's stress-free reference state.

There are many approaches to finite element simulation of the dynamics of elastic media \cite{elastodynamics};
we make use of an elegant Hamiltonian approach developed by Broughton et al \cite{bernstein, lumped_mass}, generalizing it to three dimensions and the case of large rotations. We write the Hamiltonian of an isotropic elastic solid as:

\begin{equation}
{H_{elastic} }=\sum_p V_{p}\frac{1}{2}C_{ijkl}\varepsilon^{p}_{ij}\varepsilon^{p}_{kl}+\sum_i\frac{1}{2}m_{i}v^{2}_{i}
.\label{eq:1}
\end{equation}

\noindent Here the first term represents elastic strain energy, with $p$ summing over volume elements. $V_{p}$ is the volume of element $p$ in the reference state. For an isotropic material the components of the elastic stiffness tensor $C_{ijkl}$ are determined from only two material parameters, namely the shear and bulk moduli \cite{marder_cm}. As an approximation, Broughton et al developed this formulation using the linear strain tensor, but we instead use the rotationally invariant Green-Lagrange strain tensor $\varepsilon_{ij}=\frac{1}{2}(u_{i,j}+u_{j,i}+u_{k,i}u_{k,j})$, where $\bf u$ is the displacement field.
We note that using the linearized strain tensor would make the Hamiltonian unphysical, 
as rotation of the sample would appear to cost energy. The second term represents kinetic energy in the lumped mass approximation \cite{lumped_mass} whereby the mass of each element is equally distributed among its
vertices, which are the nodes of the mesh. Here $i$ sums over all nodes, $m_{i}$ is the effective mass and $v_{i}$ the velocity of node $i$.

To account for the additional energy cost associated with the presence of a director field, we add to the potential energy, 

\begin{equation}
\begin{split}
{H_{nematic} } &= \sum_p [-\alpha\varepsilon^{p}_{ij}(Q^{p}_{ij}-Q^{o}_{ij}) +\beta(Q^{p}_{ij}-Q^{o}_{ij})^{2}] \\
&+\gamma \sum_{<p,q>}  ({Q^{p}_{ij}-Q^{q}_{ij}})^{2} 
  \end{split}
.\label{eq:2}
\end{equation}

\noindent The first term describes coupling between the strain and order parameter tensors using a 
form proposed by DeGennes \cite {DeGennes}. Here $Q^{o}_{ij}$ defines the nematic order 
in the element's reference state. The prefactor $\alpha$ controls the strength of this coupling, and DeGennes \cite {DeGennes} argued that it is of the same order of magnitude as the shear modulus $\mu$.  Variables $Q_{ij}$, $Q^{o}_{ij}$, and $\varepsilon_{ij}$ are all defined in the body frame, i.e. they are invariant under rotations in the target frame. See \cite{stenull:021807} for the relation between $Q_{ij}$ in the body and lab frames.
The second term describes ``cross-link memory," that is, the tendency of the nematic
director to prefer its orientation at crosslinking.
Thus there is an energy cost to rotate the director away from its reference state, with coupling strength $\beta$.
The third term is an energy penalty for spatial variations of the nematic director, similar to a Frank free energy in the single elastic constant approximation.  The summation is carried only over nearest neighbour elements in the mesh, as the typical domain size is of the order of the nematic correlation length \cite{Terentjev_polymono}. 

The strain tensor $\varepsilon_{ij}$ within each tetrahedral element is calculated in two steps. We calculate the displacement $\bf u$ of each node from the reference  state, then perform a linear interpolation of the displacement field within the volume element in the reference state. 
The resulting interpolation coefficients represent the derivatives $u_{i,j}$ needed to calculate
the components of the strain tensor. Details can be found in any introductory text on finite element methods, e.g. \cite{febook}. At this level of approximation, the strain is piecewise constant within each volume element. The effective force on each node is calculated as the negative derivative of the 
potential energy with respect to node displacement. To include internal dissipation in the system,
we add an additional force term which depends on the local velocity gradient  \cite{marderrubbersimulation}. This dissipation is isotropic in character and does not depend on the orientation of the director field.

To evolve the system forward in time, we assume the director is in
quasistatic equilibrium with the strain; that is, the time scale for director relaxation
is much faster than that for strain evolution as observed by Urayama \cite{urayama:041709}.
The first part of each step is elastodynamics: holding $Q_{ij}$ in each element constant, 
the equations of motion $f=ma$ for all node positions and velocities are integrated forward in time 
using the Velocity Verlet algorithm \cite{allen_tildesley}, with a time step of $10^{-8}$ sec.  In the second part of each step, we relax the nematic director in each element to instantaneously minimize the element's potential energy. Because the director rotates from a higher energy state to a lower energy state without picking up conjugate momentum, this is a source of anisotropic dissipation.
Thus in our model, as in real nematic elastomers, strains that rotate the director cause more energy dissipation than those applied parallel to the director \cite{terentjev:052701}.





\begin{figure}
\includegraphics[scale=0.3]{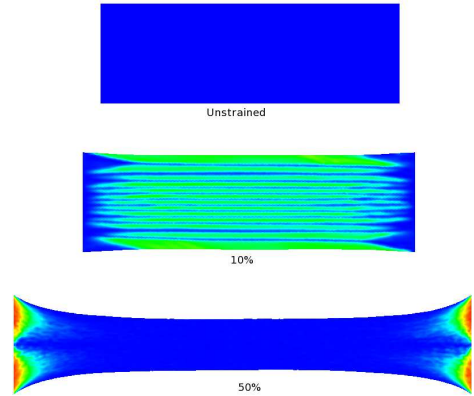}
\caption{\label{fig:1} (Color online) Simulation: stretching a nematic elastomer film at an angle of 90$^{o}$ to the director. Initially a monodomain, the director field evolves to form a striped microstructure. }
\label{fig:90degrees}
\end{figure}

We simulate uniaxial stretching in an initially monodomain nematic elastomer film of size 1.5 mm $\times$ 0.5 mm with a thickness of 50 $\mu m$, with shear modulus $\mu= 5.7 \times 10^{5} Pa$, bulk modulus $B_{r}=2.8 \times 10^{7} Pa$, and parameters $\alpha=\mu$, $\beta=0.3 \mu$, and $\gamma=10^{-7}$. We first consider the case where the director is initially oriented along the $y$ axis, transverse to the direction of applied strain. The sample is clamped on two sides and the clamped regions are constrained to move apart laterally at a constant speed of 1 mm/sec. The resulting microstructural evolution is shown in Fig.\ref{fig:90degrees}. Here color represents Jones matrix imaging of the director field as viewed through crossed polarizers parallel to the $x$ and $y$ directions; blue corresponds to a director parallel to the polarizer or analyzer, and red corresponds to a director at a $45^{o}$ angle to either. While the simulated sample is three-dimensional, the film's microstructure does not vary significantly through the thickness and can thus be visualized in 2-d.

At a strain of 8.5$\%$, the director field in the sample becomes unstable and orientational domains form, 
nucleating first from the free edges of the film. By 9$\%$ strain, the whole film is occupied by striped 
orientational domains with alternating sense of director rotation. The stripes are not uniform in width, 
being slightly larger near the free edges. Near the center of the sample, each individual stripe has a 
width of about 25 $\mu$m, which is of the same order of magnitude as that observed in 
experiment \cite{kundler_finkelmann}. This value is in reasonable agreement with the 
theoretical estimate by Warner and Terentjev \cite{Warnerbook} who predicted a stripe 
width of $h \sim \sqrt{\xi L} / \sqrt{1-1/\lambda_{1}^{3}} $; where $\xi$ is the nematic 
penetration length, $L$ is the sample width, and $\lambda_{1}$ is the strain threshold 
of the instability. The stripes coarsen as the elongation increases. Eventually this 
microstructure evolves into a more disordered state with stripes at multiple orientations. 
By reaching 35$\%$ strain, the stripes have vanished and the film is again in a monodomain state with the director oriented with the direction of strain. Only the regions near the clamped edges do not fully realign, in agreement with experimental observations \cite{kundler_finkelmann} and with the simulation studies of DeSimone \cite{desimone}. We will explore the dependence of stripe width on aspect ratio and other parameters in future work.
 
\begin{figure}
\includegraphics[scale=0.6]{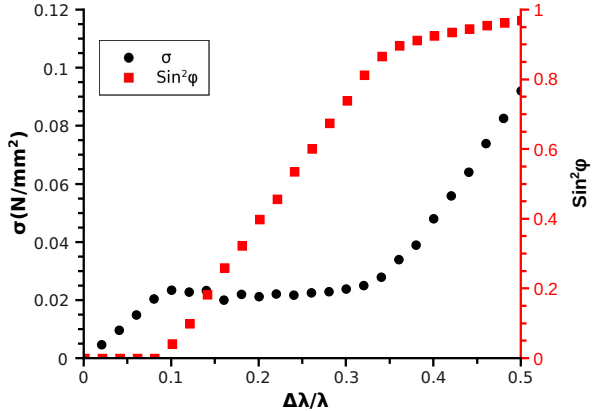}
\caption{\label{fig:2} (Color online) Engineering stress (circles) and director rotation (squares)
vs applied strain, for the system shown in Fig.~{90degrees}. Onset of director rotation and the stress-strain plateau both occur at the same strain.  }
\label{fig:ninetydeg}
\end{figure}

The resulting stress-strain response is semi-soft \cite{stenull:021807} in character, as shown in Fig.\ref{fig:ninetydeg}. The initial elastic response is linear, followed by an extended plateau running from about 8.5$\%$ to over 30$\%$  strain, after which there is a second linear regime. We also measure the average director rotation $<\sin^{2}(\phi)>$ and observe that the thresholds for both the stress-strain plateau and the rotation of the nematic director occur at the same strain. This finding demonstrates, in agreement with theory \cite{Warnerbook,stenull:021807}, that the reorientation of the system's internal degree of freedom--namely the nematic director--reduces the energy cost of the deformation. 

\begin{figure}
\includegraphics[scale=0.6]{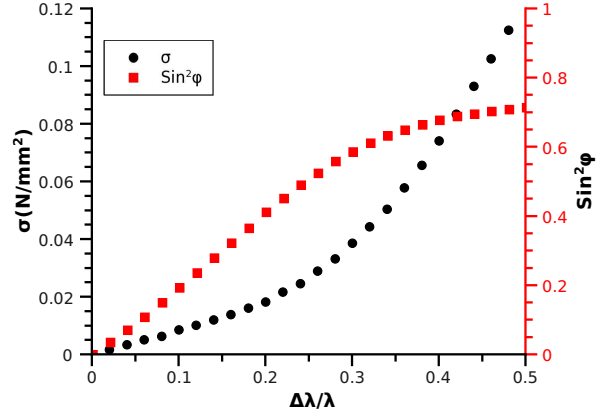}
\caption{\label{fig:3} (Color online) Engineering stress (circles) and director rotation (squares) 
vs applied strain, applied at an angle of $60^{o}$ from the nematic director. } 
\label{fig:sixtydeg} 
\end{figure}

\begin{figure}
\includegraphics[scale=0.3]{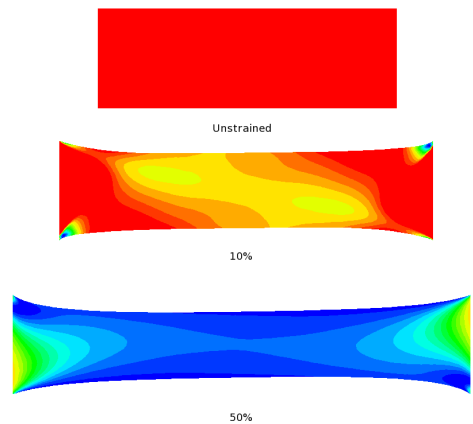}
\caption{\label{fig:4} (Color online)Simulation: stretching a nematic elastomer film at an angle of 60$^{o}$ to the director. Initially a monodomain, the director field rotates smoothly without sharp gradients in orientation. }
\label{fig:60degrees}
\end{figure}

We also performed simulations for monodomain nematic elastomer films with the initial director orientation at different angles to the pulling direction. In Fig.\ref{fig:sixtydeg} we plot the film's stress-strain response when strain is applied at an angle of 60$^{o}$ from the nematic director, which shows no plateau, and likewise director rotation shows no threshold behavior. As shown in  Fig.\ref{fig:60degrees}, the director rotates to align with the strain direction without forming stripes.  We performed additional simulations with the director at angles of 70$^{o}$ and 80$^{o}$ to the pulling direction and again found no stripe formation and no plateau in the stress-strain response. 

\begin{figure}
\includegraphics[scale=0.5]{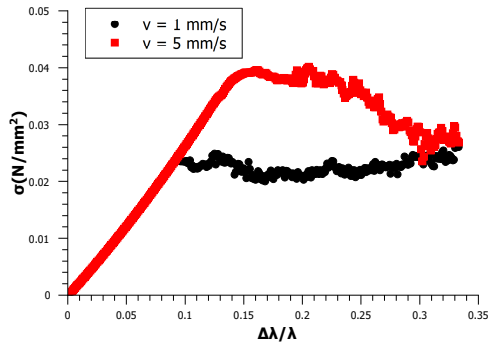}%
\caption{\label{fig:5} (Color online) Dependence of the stress-strain response on strain rate.} 
\label{fig:strainrate} 
\end{figure}

We also tried varying the applied strain rate. Fig.\ref{fig:strainrate} compares the stress-strain response for samples strained at 1 mm/sec and 5 mm/sec. The higher strain rate produces a significant stress overshoot, and stripe formation occurs at a strain of 15$\%$. This finding suggests that the threshold strain for the instability depends in a significant way on strain rate. 


\begin{figure}
\includegraphics[scale=0.24]{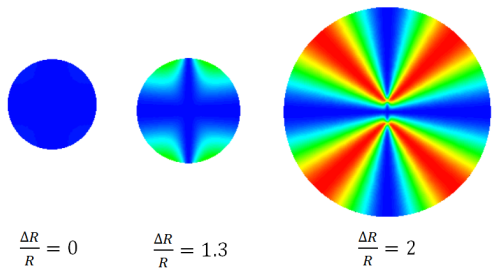}%
\caption{\label{fig:6} Simulation: A nematic elastomer disk is stretched radially. The director field smoothly transforms from a homogeneous monodomain to a radial configuration. } 
\label{fig:disk} 
\end{figure}

Next we simulated a monodomain nematic elastomer film under isotropic strain. A circular sample of diameter  1 cm and thickness 100 $\mu m$, with the director initially along the $y$ axis, was stretched radially in all directions by pulling the edge outward at constant speed.    Fig.\ref{fig:disk} shows the film at different stages of its extension, demonstrating that the director field smoothly changes from a monodomain to a radial configuration. With a careful choice of the sample's thickness, this deformed circular sheet of nematic elastomer could be used as a tunable spatial polarization converter as described in \cite{Shin-Tson}.

The simulations presented here were performed at far higher strain rates, e.g. 50$\%$ per second, than those used in typical experiments \cite{kundler_finkelmann,Clarke_terentjev_relax} where the material is allowed to relax for minutes or hours between strain increments. 
In future work we plan to apply our model to examine deformation of nematic elastomers at slower strain rates and as a function of sample geometry. 
We will also examine the role of initial microstructure and thermomechanical history in determining mechanical response. 
Using the same finite element approach, we can also test the predictions of other proposed constitutive models, and model geometries of interest for potential applications. 
Through this approach we hope to bridge the divide between fundamental theory of these fascinating materials and engineering design of devices.


\begin{acknowledgments}
We thank Profs. T. Lubensky, M. Warner, E. Terentjev and A. Desimone for fruitful discussions.
This work is supported by NSF DMR-0605889, daytaOhio, and the Ohio Supercomputer Center. FY also acknowledges the Institute for Complex Adaptive Matter - Branches Cost Sharing Fund for a postdoctoral fellowship.
\end{acknowledgments}

\bibliography{stripes}

\end{document}